# INFORMATION THEORETIC APPROACH TO ROBUST MULTI-BERNOULLI SENSOR CONTROL

*Amirali K. Gostar*   *Reza Hoseinnezhad*   *Alireza Bab-Hadiashar*

School of Aerospace, Mechanical & Manufacturing Engineering
RMIT University
Victoria 3083, Australia

**ABSTRACT**

A novel sensor control solution is presented, formulated within a Multi-Bernoulli-based multi-target tracking framework. The proposed method is especially designed for the general multi-target tracking case, where no prior knowledge of the clutter distribution or the probability of detection profile are available. In an information theoretic approach, our method makes use of Rènyi divergence as the reward function to be maximized for finding the optimal sensor control command at each step. We devise a Monte Carlo sampling method for computation of the reward. Simulation results demonstrate successful performance of the proposed method in a challenging scenario involving five targets maneuvering in a relatively uncertain space with unknown distance-dependent clutter rate and probability of detection.

## 1. INTRODUCTION

Sensor control, in general, comprises a *multi-object filtering* process and an *optimal decision-making* method. The sensor control problem, concerned in this paper, is focused on controlling the states of mobile sensors that are used for multi-target tracking. In any sensor control framework, the control commands are the direct outputs of the above mentioned *decision-making* component which is the process of selecting the optimal control command. Thus, most of the existing solutions are focussed on devising and improving that component, assuming that the *multi-object filtering* framework can effectively utilize the sensor measurements and return accurate estimates of the number and states of the targets. However, the multi-target tracking framework, employed with sensor control, itself plays a significant role in the overall performance of the scheme in terms of real-time accuracy and robustness.

A few sensor control solutions have been recently developed for multi-target tracking scenarios where tracking is performed by the multi-object filters based on Finite Set Statistics (FISST) theory [1–6]. Similar to classical approaches, the general form of Bayesian recursion involved in FISST-based methods is not computationally tractable [1]. Consequently, several approximations are suggested. PHD filter and its extended version, CPHD filter [1], Multi-Bernoulli filters [1, 7] and Vo-Vo prior filter [8] are the most well-known instances of such approximations.

The most common approach for the objective function in sensor control is *information-driven* in which sensor control is aimed at improving the information content of the multi-object distribution by optimizing some measure of information gain. The most common choice for an objective function in information-driven methods is Rényi divergence function. Ristic *et al.* [2] used Rényi divergence as the objective function in conjunction with random set filter and PHD-based filter for the scenarios where clutter rate and uncertainty in sensor Filed of View (FoV) are known. This paper presents a novel sensor control solution designed to work within a robust Multi-Bernoulli-based multi-target tracking framework. Our sensor control solution does not need any prior knowledge of the clutter distribution or the probability of detection profile.

## 2. ROBUST MULTI-BERNOULLI FILTER

Vo *et al.* [9] have recently tackled the problem of multi-Bernoulli filtering for cases where clutter intensity and detection probability profile are unknown. In this solution, the *detection probability* is *augmented* to the multi-target state, and propagated in time. A set of *clutter generators* is used to create hypothetical targets associated with clutter measurements. The transition and observation models for the clutter-associated targets are taken to be similar to actual targets. Those two types of targets form a *hybrid space* $\breve{\mathcal{X}} = \mathcal{X}^{(0)} \cup \mathcal{X}^{(1)}$ where $(0)$ and $(1)$ superscripts denote the space of clutter and actual targets, respectively. Their augmented multi-target state includes a state $\breve{a}$ (as the probability of detection) and a multi-Bernoulli set state $X$.

The multi-Bernoulli random set of targets is the union of an ensemble of $M$ Bernoulli sets, each having a probability of existence, $r^{(i)(u)}$, and two single-object densities denoted by $p^{(i)(u)}(\breve{a}, \mathbf{x})$ where $u = 1$ corresponds to objects that are actual targets, and $u = 0$ corresponds to clutters.

In this method, the prior multi-target distribution at time $k-1$ is used in the prediction step that incorporates models of the dynamic and birth of targets and clutter generators. The multi-target distribution is then updated using current sensor measurements—see [9] for details of prediction and update steps and formulas. In the following section, we explain how the above method can be modified to incorporate a sensor selection step.

## 3. SENSOR CONTROL FRAMEWORK

We formulate the sensor control problem in the POMDP framework in conjunction with a robust multi-Bernoulli filter [9] to concurrently solve the sensor control and multi-object estimation problems with unknown clutter intensity and sensor FoV.

POMDP is a generalized form of Markov decision process (MDP) which is suitable for sensor planning manifested by a series of control commands (actions). In POMDP framework, there is no direct access to the *states* and decisions are made using only uncertain observations. A POMDP at any time step $k$ could be defined as a tuple; $\Psi = \{X_k, \mathbb{S}, f_{k|k-1}(x_k|x_{k-1}), Z_k, g_k(z|x), \vartheta(X_{k-1}, s, X_k)\}$, where $X_k$ is a finite set of single-object states; $\mathbb{S}$ defines a set of sensor control commands; $f_{k|k-1}(x_k|x_{k-1})$ is a transition model for single-object state; $Z_k$ comprises a finite set of observations; $g_k(z|x)$ is a stochastic measurement model; and $\vartheta(X_{k-1}, s, X_k)$ is an objective function that returns a reward or cost for transition from the multi-object state $X_{k-1}$ to the state $X_k$ by applying an action command $s \in \mathbb{S}$.

The purpose of the sensor control is to find the control command $\mathring{s}$ that optimizes the objective function. In an information theoretic approach, the reward function depends on the distributions of $X_{k-1}$ and $X_k$, as well as the future measurements which are distributed according to the choice of the control command. Thus, the control command $\mathring{s}$ is commonly chosen to maximize the statistical mean of the reward function over all future measurements,

$$\mathring{s}_k = \underset{s \in \mathbb{S}}{\operatorname{argmax}} \left\{ \mathbb{E}_{Z_k} [\vartheta(X_{k-1}, s, X_k)] \right\}. \quad (1)$$

## 4. ROBUST MULTI-BERNOULLI SENSOR CONTROL

Suppose that at time $k-1$, the posterior multi-object density is approximated by multi-Bernoulli RFS. Using this distribution, the predicted multi-Bernoulli state is computed. For each sensor command the updated multi-object distribution at time $k$ is computed and denoted by $\{(r_k^{(i)}, p_k^{(i)(u)}(\cdot))\}_{i=1}^{M_k}$ where each density $p_k^{(i)(u)}(\cdot)$ is approximated by a set of support points $\{(\breve{a}_k^{(i,j)(u)}, \mathbf{x}_k^{(i,j)(u)})\}_{j=1}^{L_k^{(i)(u)}}$ with weights $\{w_k^{(i,j)(u)}\}_{j=1}^{L_k^{(i)(u)}}$. Following [2], we implement Rényi divergence as the objective function. We note that the value of the objective function depends on future measurements which have stochastic variations even for a determined control command. Following [10], we skip the substantial computations needed for averaging over all future measurements, and use the *predicted ideal measurement set* (PIMS) approach to generate the measurements required to update the predicted distribution in the decision making step.

The PIMS is produced as follows. The cardinality and state of the multi-object RFS are estimated from the predicted multi-Bernoulli distribution. These will include actual target estimates and points estimates corresponding to generated clutters. Using the estimated number and states of the objects, a set of noise-free measurements are created and denoted by $Z$. This set would include clutter measurements associated with clutter generators that are not discriminated from real objects in the prediction step.[1] Thus, it is necessary to perform an initial update step, using $Z$ as the measurement set. The number of target objects and their states are then estimated from this primarily updated distribution. Using the estimated target states, for each control command, $s$, a single noise-free measurement is generated for each single object, and the set of such measurements is taken as the PIMS, denoted by $\mathring{Z}(s)$. For each control command, $s$, its corresponding PIMS, $\mathring{Z}(s)$, is then used to update the multi-Bernoulli multi-object density from which the objective function is calculated and the optimum control command is selected accordingly.

In [2], Rényi divergence was employed as the generalization of the Kullback-Leibler divergence. Rényi divergence measures information gain between two densities:

$$I_\alpha(f_1, f_2) = \frac{1}{\alpha-1} \log \int [f_1(X)]^\alpha [f_2(X)]^{1-\alpha} \delta X. \quad (2)$$

In a target tracking context, the first distribution is usually the updated distribution, $\pi_{k+1}(X_{k+1}|Z_{1:k}, u_{0:k-1}, Z_{k+1}, u_k)$, and the second distribution is the prediction distribution, $\pi_{k+1|k}(X_{k+1}|Z_{1:k}, u_{0:k-1})$. The $\alpha$ parameter tunes the emphasis on each of the two distributions. Similar to [11], our sensor control method is based on computing the reward function for each control command, in which the updated multi-object distribution is the one updated using the PIMS. However, in the particle approximation of the reward function derived in [11],

$$\mathcal{D}(\mathbf{s}_k) \approx \frac{1}{\alpha-1} \log \frac{\sum_{i=1}^n w_k^i [g_{k+1}(\mathbf{Z}_{k+1}|\mathbf{X}_{k+1|k}^i, \mathbf{s}_k)]^\alpha}{\left[\sum_{i=1}^n w_k^i [g_{k+1}(\mathbf{Z}_{k+1}|\mathbf{X}_{k+1|k}^i, \mathbf{s}_k)]\right]^\alpha},$$

computation of the likelihood function $g_{k+1}(\mathbf{Z}_{k+1}|\mathbf{X}_{k+1|k}^i, \mathbf{s}_k)$ needs the knowledge of clutter intensity and probability of

---
[1]Referring to [9], the predicted existence probabilities are not separately computed for $u = 0, 1$ (see equation (11) in [9]). This is while the updated $r^{(i)}$'s and $p^{(i)}(\cdot)$'s are separately calculated for $u = 0, 1$—see equations (14)-(19) in [9].

**Algorithm 1** Monte Carlo sampling of a multi-Bernoulli distribution with given parameters and particles.

    INPUTS: probabilities of existence $\mathbf{r} = [r_1 \cdots r_M]^\top$, particles matrix $P_{M \times L_{\max}} = [x_{ij}]$, and number of Monte Carlo samples $L$.
    OUTPUTS: $L$ sets, each being a Monte Carlo sample of the multi-Bernoulli distribution, in the form of $X_\ell = \{x_{\ell,1}, \ldots, x_{\ell,n_\ell}\}$, where $n_\ell \leq M$ is the cardinality of the $\ell$-th set.
1: From the size of the particles matrix, find $M$ and $L_{\max}$.
2: **for** $\ell = 1, L$ **do**
3:     $X_\ell \leftarrow \varnothing$
4:     **for** $i = 1, M$ **do**
5:         Generate $u \sim U(0, 1)$.
6:         **if** $u < r_i$ **then**
7:             Generate $v \sim U(0, 1)$.
8:             $j \leftarrow \lceil L_{\max} v \rceil$.    ▷ A random index between 1 and $L_{\max}$.
9:             $X_\ell \leftarrow X_\ell \cup \{x_{ij}\}$.
10:        **end if**
11:     **end for**
12: **end for**

detection profiles—see equation (20) in [11]. In the absence of such information, we propose to directly compute the reward function by Monte Carlo sampling of the updated multi-object multi-Bernoulli distribution.

Monte Carlo sampling of a general multi-Bernoulli distribution can be performed as follows. Assume that a multi-Bernoulli distribution is given with parameters $\{r_i, p_i(\cdot)\}_{i=1}^M$ with particle approximation $p_i(x) \approx \sum_{j=1}^{L_i} w_{ij} \, \delta(x - x_{ij})$. Without loss of generality, we assume that after resampling, $L_{\max}$ particles with equal weights are created for each Bernoulli component. Algorithm 1 shows our proposed pseudocode to generate $L$ Monte Carlo samples of the above described multi-Bernoulli distribution.

Let us denote the ensemble of Monte Carlo samples of the updated distribution by $\{X_\ell\}_{\ell=1}^L$ where each sample is itself a set of particles $X_\ell = \{x_{\ell,1}, \ldots, x_{\ell,n_\ell}\}$, in which $n_\ell \leq M$ is the cardinality of the $\ell$-th sample. Particle approximation of the updated distribution is given by the following linear combination of multi-object Dirac delta functions:

$$\pi_{k+1}(X|\mathring{Z}(u_k)) \approx \sum_{\ell=1}^L \frac{1}{L} \delta(X - X_\ell). \quad (3)$$

For a sufficiently large number of Monte Carlo samples, we have:

$$\forall h(\cdot), \int h(X) \pi_{k+1}(X|\mathring{Z}(u_k)) \, \delta X = \sum_{\ell=1}^L \frac{1}{L} h(X_\ell) + O(L^{-1}). \quad (4)$$

Substituting the arbitrary function $h(X)$ in (4) with $\left[\frac{\pi_{k+1|k}(X)}{\pi_{k+1}(X|\mathring{Z}(u_k))}\right]^{1-\alpha}$ would lead to:

$$\frac{1}{\alpha-1} \log \int \left[\pi_{k+1}(X|\mathring{Z}(s_k))\right]^\alpha \left[\pi_{k+1|k}(X)\right]^{1-\alpha} \delta X$$
$$= \sum_{\ell=1}^L \frac{1}{L} \left[\pi_{k+1|k}(X_\ell) / \pi_{k+1}(X_\ell|\mathring{Z}(s_k))\right]^{1-\alpha} + O(L^{-1}). \quad (5)$$

We note that the left side of the above equation is same as the reward function, $\mathcal{D}(s_k)$, and thus it can be computed by discarding the $O(L^{-1})$ term. The multi-Bernoulli distribution terms can be directly calculated. For general multi-Bernoulli parameters $\{r^{(i)}, p^{(i)}(\cdot)\}_{i=1}^M$, the density at $X = X_\ell$ is given by [1] p. 369:

$$\pi(\varnothing) = \prod_{i=1}^M (1 - r^{(i)}). \text{ if } |X_\ell| > M, \, \pi(X_\ell) = 0, \text{ otherwise}$$
$$\pi(X_\ell) = \pi(\varnothing) \sum_{1 \leq i_1 \neq \ldots \neq i_{|X_\ell|} \leq M} \prod_{j=1}^{|X_\ell|} \frac{r^{(i_j)} p^{(i_j)}(x_{\ell,j})}{1 - r^{(i_j)}}. \quad (6)$$

In order to compute the density terms $p^{i_j}(\cdot)$ in (6), we note that for the updated multi-Bernoulli density, the Monte Carlo samples are drawn using the particles of the updated single Bernoulli components—see Algorithm 1. More precisely, when computing the updated multi-Bernoulli density $\pi_{k+1}(X_\ell)$, the argument of each of the $p^{i_j}(\cdot)$ terms in (6), $x_{\ell,j}$, definitely coincides with one of the particles representing the updated $i_j$-th single-Bernoulli distribution $p_{k+1}^{(i_j)}(\cdot)$. This particle is found and its weight is used for the density term in (6). When computing the predicted density $\pi_{k+1|k}(X_\ell)$, the density term $p^{(i_j)}(x_{\ell,j})$ is computed by kernel density estimation using Gaussian kernels centered at the predicted particles for the $i_j$-th Bernoulli component.

## 5. SIMULATION RESULTS

A challenging non-linear multi-target tracking scenario, similar to the one reported in [5], is employed to evaluate the performance of the proposed robust multi-Bernoulli sensor control method. In this scenario, a mobile sensor is employed to manoeuvre in a dynamic environment of size 1000m × 1000m. The sensor regularly scans the surveillance area and returns a set of bearing and range measurements corresponding to detected targets, each in the form of $z_k = [\theta_k, \mathfrak{R}_k]^\top$. A total of five targets appear in the scene and manoeuvre in the surveillance area. At each time step, $k$, the sensor location is denoted by $s_k = [x_{s_k} \, y_{s_k}]^\top$. The sensor enters the surveillance area at position (10m, 10m).

We ran the proposed robust multi-Bernoulli sensor control method to estimate the number and locations of the targets for a sequence of 35 steps. Intuitively, the sensor control method is expected to select those sensor positions that are closer to existing targets and end up in the vicinity of them. The rationale behind this expectation is that the noise power for range measurements increases with the distance and the detection probability decreases for large distances—see equations (8)-(9) in [5].

Fig. 1 shows average estimates of cardinality and clutter intensity calculated by 200 Monte Carlo runs of the proposed method. It is observed that the clutter intensity estimates shown in Fig. 1(a) gradually approach the ground-truth value of 10 and fluctuate around it with a relatively small standard deviation.

For the averaged cardinality estimates shown in Fig. 1(b),

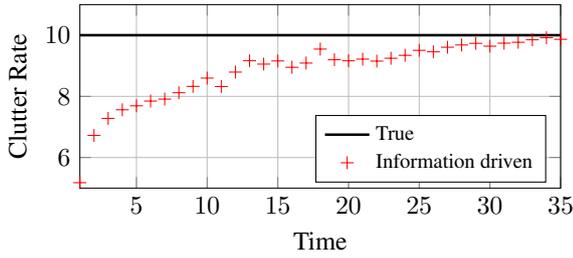

(a) Comparision of the statistical mean of clutter intesity estimates for information driven methods.

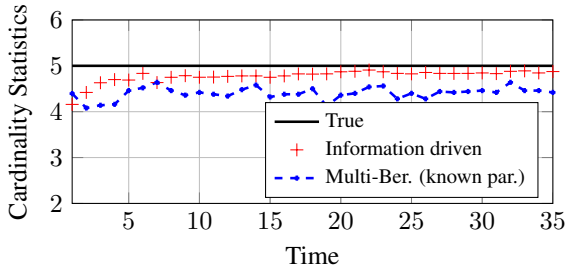

(b) Comparision of the statistical mean of cardinality estimates of the information driven methodes as well as multi-Bernoulli method with known parameters [3].

**Fig. 1**. Clutter intensity and cardinality estimates averaged over 200 MC runs.

we have also included the results returned by sensor control using multi-Bernoulli filtering with known clutter intensity and probability of detection as reported in [3]. In that method, we used an overestimated value of 15 for the clutter intensity and a fixed value of 0.98 for probability of detection, which contrasts to the ground-truth probability of detection that decreases with sensor-target distance. We observe that in terms of estimating the cardinality, the proposed method leads to better accuracy, as the averaged cardinality estimates are closer to the ground-truth number of 5 targets.

## 6. CONCLUSION

A novel approach to sensor control in multi-target tracking application with unknown sensor FoV and clutter was presented. The proposed method uses Rényi divergence as the reward function, and computation of this reward is suggested to be performed using Monte Carlo sampling of predicted and updated multi-object densities. Simulation results demonstrate satisfactory performance of the sensor control method in estimating the number and states of five targets in a challenging scenario. It was also shown to improve the estimation accuracy in comparison with the common solutions in which the sensor FoV parameters are assumed known but the values used are practically deviated from the ground-truth. The superior performance of this method mainly lays in its *adaptive* nature. Indeed, in the absence of accurate knowledge of the measurement process, inaccurate assumptions for measurement process parameters would lead to inaccurate results and frequently missed targets. This is while without the need for such prior information, our method intrinsically adapts the multi-target filtering process to work best with the measurements received from the selected sensor command.

## Acknowledgment

This work was supported by the ARC Discovery Project grant DP130104404.